\documentclass[final,11pt]{article}
\usepackage{multirow}
\usepackage[nottoc]{tocbibind}
\usepackage{geometry}
\usepackage{subcaption}
\usepackage{placeins}
\usepackage[affil-it, auth-sc]{authblk}
\usepackage[english]{babel}
\usepackage[utf8]{inputenc}
\usepackage{graphicx}
\usepackage{amssymb}
\usepackage{amsthm}
\usepackage{amsmath}
\usepackage{amsfonts}
\usepackage{verbatim}
\usepackage{lineno}
\usepackage{todonotes}
\presetkeys{todonotes}{inline, color=blue!30, size=\small}{}

\usepackage{color, soul}
\definecolor{darkred}{rgb}{0.9, 0.0, 0.0}
\definecolor{darkgreen}{rgb}{0.0, 0.5, 0.0}

\usepackage[colorlinks,bookmarks,linkcolor=darkred, citecolor=darkgreen]{hyperref}
\usepackage{eso-pic}
\usepackage[sort&compress,numbers]{natbib}
\usepackage{doi}
\usepackage{paralist,epsfig}
\usepackage{relsize}
\usepackage{nicefrac,esint}
\usepackage{float}
\usepackage{cleveref}

\begin{document}

\AddToShipoutPictureFG*{
    \AtPageUpperLeft{\put(-60,-40){\makebox[\paperwidth][r]{LA-UR-22-32611}}}  
    }
    
\AddToShipoutPictureFG*{
    \AtPageUpperLeft{\put(-60,-60){\makebox[\paperwidth][r]{FERMILAB-PUB-23-420-T}}}  
    }
    
\title{Nucleon axial-vector form factor and radius  \\ from future neutrino experiments}
\author[1]{Roberto~Petti \thanks{Roberto.Petti@cern.ch}}
\affil[1]{University of South Carolina, Department of Physics and Astronomy, Columbia, SC 29208 USA \vspace{1.2mm}}
\author[2,3]{Richard~J.~Hill \thanks{Richard.Hill@uky.edu}}
\affil[2]{University of Kentucky, Department of Physics and Astronomy, Lexington, KY 40506 USA \vspace{1.2mm}}
\affil[3]{Fermilab, Theoretical Physics Department, Batavia, IL 60510 USA \vspace{1.2mm}}
\author[4]{Oleksandr~Tomalak \thanks{tomalak@lanl.gov}}
\affil[4]{Theoretical Division, Los Alamos National Laboratory, Los Alamos, NM 87545 USA \vspace{1.2mm}}

\date{\today}

\maketitle

\begin{abstract}
Precision measurements of antineutrino elastic scattering on hydrogen from future neutrino experiments offer a unique opportunity to access the low-energy structure of protons and neutrons. We discuss the determination of the nucleon axial-vector form factor and radius from antineutrino interactions on hydrogen that can be collected at the future Long-Baseline Neutrino Facility, and study the sources of theoretical and experimental uncertainties. The projected accuracy would improve existing measurements by $1$ order of magnitude and be competitive with contemporary lattice-QCD determinations, potentially helping to resolve the corresponding tension with measurements from (anti)neutrino elastic scattering on deuterium. We find that the current knowledge of the nucleon vector form factors could be one of the dominant sources of uncertainty. We also evaluate the constraints that can be simultaneously obtained on the absolute $\bar \nu_\mu$ flux normalization.
\end{abstract}

\section{Introduction}

Elastic lepton-nucleon scattering probes the distribution of charge, magnetization, spin, and isospin within protons and neutrons, via the corresponding form factors. The combined information obtained from electron, muon, and (anti)neutrino scattering data can provide fundamental insights to understand the hadronic structure of free nucleons. Such studies are a required input for the modeling of lepton-nucleus interactions, as well as for our understanding of modifications to nucleon properties within nuclei.

In the approximation of one exchanged boson, elastic lepton-nucleon scattering is fully characterized by the nucleon form factors. Electromagnetic interactions are described by two form factors, Dirac and Pauli or, equivalently, electric and magnetic, which can be extracted from experimental data~\cite{Dombey:1969wk,Akhiezer:1974em,Bernauer:2010wm,Bernauer:2013tpr,Xiong:2019umf,Punjabi:2015bba,Ganichot:1972mb,Bosted:1989hy,Perdrisat:2006hj,Jones:1999rz,Gayou:2001qd,Punjabi:2005wq,Puckett:2010ac,Ron:2011rd,Zhan:2011ji}, from model calculations~\cite{Bernard:2001rs,Bernard:1998gv,Schindler:2006it,Schindler:2006jq,Ando:2006xy,Scherer:2009bt,Yao:2017fym,Alarcon:2017lhg,Alarcon:2017ivh,Zhang:2019iyx,Chung:1991st,Cardarelli:1995dc,Brodsky:1997de, Miller:2002ig, Miller:2002qb,Ma:2002ir,Ma:2002xu,Pasquini:2007iz,Cloet:2012cy,Brodsky:2014yha,Mamedov:2022yyd}, or from matrix elements of quark currents on the lattice~\cite{Gockeler:2003ay,Alexandrou:2018sjm,Yamazaki:2009zq,Chambers:2017tuf,Capitani:2015sba,Bruno:2014jqa,Bratt:2010jn,Kronfeld:2019nfb,Ishikawa:2018rew,Hasan:2019noy,RQCD:2019jai,Jang:2019vkm,Bali:2019yiy,Alexandrou:2020okk,Park:2021ypf,Djukanovic:2022wru,Jang:2023zts}. The charged-current elastic neutrino-neutron and antineutrino-proton processes require two additional form factors: axial-vector and induced pseudoscalar. In modern and future neutrino experiments~\cite{Nunokawa:2007qh,NOvA:2007rmc,T2K:2011qtm,KamLAND:2013rgu,MicroBooNE:2015bmn,JUNO:2015zny,Hyper-KamiokandeProto-:2015xww,McConkey:2017dsv,RENO:2018dro,DayaBay:2018yms,Machado:2019oxb,MicroBooNE:2019nio,DoubleChooz:2019qbj,Farnese:2019xgw,T2K:2019bcf,NOvA:2019cyt,DUNE:2020ypp,JUNO:2021vlw,NuSTEC:2017hzk,Carlson:2014vla,Benhar:2015wva,Rocco:2015cil,Pastore:2017uwc,Lynn:2019rdt}, the contribution from the induced pseudoscalar form factor is suppressed by the charged lepton mass for electron and muon flavors and can be well approximated by the partially conserved axial current ansatz in the assumption of pion pole dominance~\cite{LlewellynSmith:1971uhs,Fuchs:2002zz,Kaiser:2003dr,Lutz:2020dfi,Chen:2020wuq}. Conversely, the axial-vector form factor provides substantial contributions to (anti)neutrino scattering cross sections, making it a critical element for precision measurements of (anti)neutrino-nucleon interactions and neutrino oscillation parameters in long-baseline experiments.

Precision measurements of antineutrino-proton elastic scattering in hydrogen can concurrently offer a valuable tool to address some of the limitations of high-energy neutrino scattering experiments using nuclear targets~\cite{Petti:2019asx,Petti:2022bzt}. Events with small energy transfer allow the determination of the shape of the $\bar \nu_\mu$ flux as a function of energy with an accuracy of $\sim$1\% in conventional wideband beams, while events at small momentum transfer $Q^2 \to 0$ can be used to extract the absolute $\bar \nu_\mu$ flux normalization~\cite{Duyang:2019prb}. Elastic antineutrino-proton interactions can help to calibrate the reconstructed antineutrino energy and to constrain the corresponding systematic uncertainties arising from nuclear smearing, via a comparison with similar interactions from nuclear targets~\cite{Petti:2022bzt,Petti:2023osk}. Given the absence of a physical neutron target and the sizable nuclear effects in the deuteron, antineutrino-proton scattering is also a valuable probe of neutrino-neutron elastic processes via isospin symmetry.

Existing measurements of the nucleon axial-vector form factor from (anti)neutrino elastic scattering are scarce. Most determinations were obtained by bubble chamber experiments on a deuterium target~\cite{Mann:1973pr,Barish:1977qk,Miller:1982qi,Baker:1981su,Kitagaki:1983px} and suffer from the limited statistics and a somewhat inconsistent treatment of systematic uncertainties. Tensions have been observed between deuterium measurements and several lattice-QCD determinations of the nucleon axial-vector form factor~\cite{Meyer:2022mix}. We note that both experimental measurements and nuclear models indicate that bound nucleons are modified by the nuclear environment in the deuteron~\cite{Griffioen:2015hxa,Singh:1971md,Shen:2012xz,Kulagin:2007ju,Kulagin:2014vsa,Alekhin:2022tip,Alekhin:2022uwc,Accardi:2016qay} and such modifications play an important role in the form factor extraction from the deuterium data~\cite{Meyer:2016oeg}. A recent measurement was obtained from antineutrino scattering on the hydrogen atoms within a plastic scintillator (CH) target~\cite{MINERvA:2023avz}, with a statistics significantly higher than the bubble chamber data. The total uncertainty of this measurement is too large to resolve the observed differences between deuterium and lattice-QCD fits~\cite{Tomalak:2023pdi}.

Future measurements of (anti)neutrino interactions on free protons using the ``solid" hydrogen technique~\cite{Petti:2019asx,Petti:2022bzt,Duyang:2018lpe} can provide high-statistics samples with commensurate systematics. The interactions on free protons are obtained from a model-independent subtraction between measurements on multiple thin --- about 1.5\% of radiation length --- graphite (C) and polypropylene (CH$_2$) targets alternated within a low-density detector ($\leq$0.18 g/cm$^3$) allowing an accurate characterization of the various event topologies. The selection of antineutrino-hydrogen elastic interactions was studied~\cite{Duyang:2018lpe,Duyang:2019prb} in the context of the intense beam planned at the Long-Baseline Neutrino Facility (LBNF)~\cite{Abi:2020evt,Rout:2020cxi}. In this work, we investigate the corresponding physics opportunities for the determination of the nucleon axial-vector form factor and the axial-vector radius. In particular, we discuss the expected sources of uncertainty and compare the results with the existing measurements and theoretical predictions. We also evaluate the constraints that can be simultaneously obtained on the absolute $\bar \nu_\mu$ flux normalization.

\section{Analysis Framework \label{sec:framework}}

In order to study the uncertainties in the extraction of the nucleon axial-vector form factor $F_A$ and axial-vector radius $r_A$, we generate the cross section for antineutrino elastic scattering events on hydrogen with the vector form factors (VFFs) from Ref.~\cite{Borah:2020gte}\footnote{We use the default fit ``iso ($1\,{\rm GeV}^2$)" from Ref.~\cite{Borah:2020gte}.  Similar results are obtained using an alternate fit including higher-$Q^2$ data ``iso ($3\,{\rm GeV}^2$)".} and the axial-vector form factor from Ref.~\cite{Meyer:2016oeg}. Recall the well-known decomposition~\cite{LlewellynSmith:1971uhs}:
\begin{equation}
\frac{\mathrm{d} \sigma}{\mathrm{d} Q^2} \left( E_\nu, Q^2 \right) = \frac{\mathrm{G}_\mathrm{F}^2 |V_{ud}|^2}{2\pi} \frac{M^2}{E_\nu^2} \left[ \left( \tau + r_\ell^2 \right)A(\nu,~Q^2) + \frac{\nu}{M^2} B(\nu,~Q^2) + \frac{\nu^2}{M^4} \frac{C(\nu,~Q^2)}{1+ \tau} \right] \,, \label{eq:xsection_CCQE}
\end{equation}
where $\nu = E_\nu / M - \tau - r_\ell^2$, $\tau = Q^2/(4M^2)$, $r_\ell = m_\ell/(2M)$, $M$ is the nucleon mass, $m_\ell$ is the charged lepton mass, and $E_\nu$ is the neutrino energy. The multiplicative factors are the Fermi coupling constant $\mathrm{G}_\mathrm{F}$ and the Cabibbo-Kobayashi-Maskawa matrix element $V_{ud}$. In the approximation of one exchanged boson, the structure-dependent parameters $A$, $B$, and $C$ are expressed in terms of the nucleon electric $G^V_E$, magnetic $G^V_M$, axial-vector $F_A$, and pseudoscalar $F_P$ form factors as
\begin{align}
A &= \tau  \left(G^V_M\right)^2 - \left(G^V_E\right)^2 + (1+ \tau) F^2_A - r_\ell^2 \left( \left(G^V_M\right)^2 + \left( F_A + 2 F_P \right)^2 - 4 \left( 1 + \tau \right) F^2_P \right) \,, \\
B &=  4 \tau F_A G^V_M \,,  \\
C &= \tau \left(G^V_M\right)^2 + \left(G^V_E\right)^2 + (1 + \tau) F^2_A.
\end{align}
We evaluate the corresponding cross section in bins of the momentum transfer $Q^2 = - \left( k_{\bar{\nu}_\mu} - k_{\mu^+} \right)^2$, with $k_{\bar{\nu}_\mu}$ and $k_{\mu^+}$ being the momenta of the incoming antineutrino and the outgoing muon, respectively.

Predicted event rates are obtained by integrating the cross section in Eq.~(\ref{eq:xsection_CCQE}) over the antineutrino flux spectrum expected at the LBNF from Ref.~\cite{dune_page}. Following Ref.~\cite{Duyang:2019prb}, we use 30 bins with a bin size of 0.05 GeV$^2$ up to 1.5 GeV$^2$ and for each bin we assign the corresponding projected statistical and systematic uncertainties. The latter include the effect of the $\bar \nu_\mu$ flux shape, momentum scale, and angle reconstruction. As default, we consider a total exposure equivalent to $5.5 \times 10^{21}$ protons on target (POT), with a detector based on a straw tube tracker and an integrated solid hydrogen target with a fiducial H mass of about 700 kg. A similar detector will be part of the near detector complex of the Deep Underground Neutrino Experiment (DUNE)~\cite{DUNE:2022aul}.

We perform a fit to the generated $Q^2$ distributions by fixing the nucleon vector form factors to the values from Ref.~\cite{Borah:2020gte}. The axial-vector form factor is expressed in terms of a $z$-expansion as
\begin{align}
    F_A \left( Q^2 \right) = \sum \limits_{k=0}^{k_\mathrm{max}} a_k z \left( Q^2 \right)^k, \qquad z \left( Q^2 \right) = \frac{\sqrt{t_\mathrm{cut} + Q^2} - \sqrt{t_\mathrm{cut} - t_0}}{\sqrt{t_\mathrm{cut} + Q^2} + \sqrt{t_\mathrm{cut} - t_0}}, \label{eq:zexp}
\end{align}
with the parameters $t_0 = - 0.28~\mathrm{GeV}^2$ and $t_\mathrm{cut} = 9 m_{\pi}^2 = 0.1764~\mathrm{GeV}^2$. The normalization is defined by $g_A = \sum \limits_{k=0}^{k_\mathrm{max}} a_k z \left( 0 \right)^k$. Perturbative QCD behavior at large $Q^2$~\cite{Lepage:1980fj,Chernyak:1983ej} (i.e., $F_A \sim 1/Q^4$ up to logarithms) implies four sum rules~\cite{Becher:2005bg,Lee:2015jqa,Meyer:2016oeg},
\begin{align}
    \sum \limits_{k=n}^{k_\mathrm{max}} k \left( k - 1 \right) ... \left( k - n + 1 \right)  a_k = 0, \qquad n = 0, 1, 2, 3.
\end{align}
We consider initially only the information from the shape of the $Q^2$ distribution, by fixing the corresponding integral to unity in the fit. The normalization of the axial-vector form factor is set to $g_A = -1.2723$~\cite{Meyer:2016oeg}. We employ Gaussian bounds on all $z$ expansion coefficients: $|a_k|_\mathrm{max} = 5 |a_0|$ for $k = 0 \dots k_\mathrm{max}$. As a default, we truncate the $z$-expansion of the nucleon axial-vector form factor after $k_\mathrm{max} = 8$ and vary the four free parameters $a_1,~a_2,~a_3,~a_4$, following the common practice of Refs.~\cite{Meyer:2016oeg,Borah:2020gte}. This choice in the number of free parameters is motivated by the observation that $|z|^5$ is bounded by $0.31\%$ in the range $0\le Q^2 \le 1.5\,{\rm GeV}^2$, which is below the expected experimental accuracy of the data we are considering. For comparison, we also truncate the $z$-expansion after $k_\mathrm{max} = 7$ (three free parameters) and $k_\mathrm{max} = 9$ (five free parameters). The axial-vector radius $r_A$ is given by the conventional definition:
\begin{align}
    r_A^2  \; = \;  {- \frac{6}{F_A \left( 0 \right)} \frac{\mathrm{d} F_A \left( Q^2 \right)}{ \mathrm{d} Q^2} \Bigg |_{Q^2 = 0}}.
\end{align}

\section{Results and Discussion}

We begin by evaluating the achievable precision for the axial-vector form factor and radius in a fit to the shape of the normalized and flux-averaged $Q^2$ distribution. We then consider an enlarged fit to simultaneously constrain the absolute $\bar \nu_\mu$ flux and the axial-vector form factor.

\subsection{Shape-only fits} \label{sec:shape_only}

\begin{figure}[]
    \centering
    \includegraphics[height=0.4\textwidth]{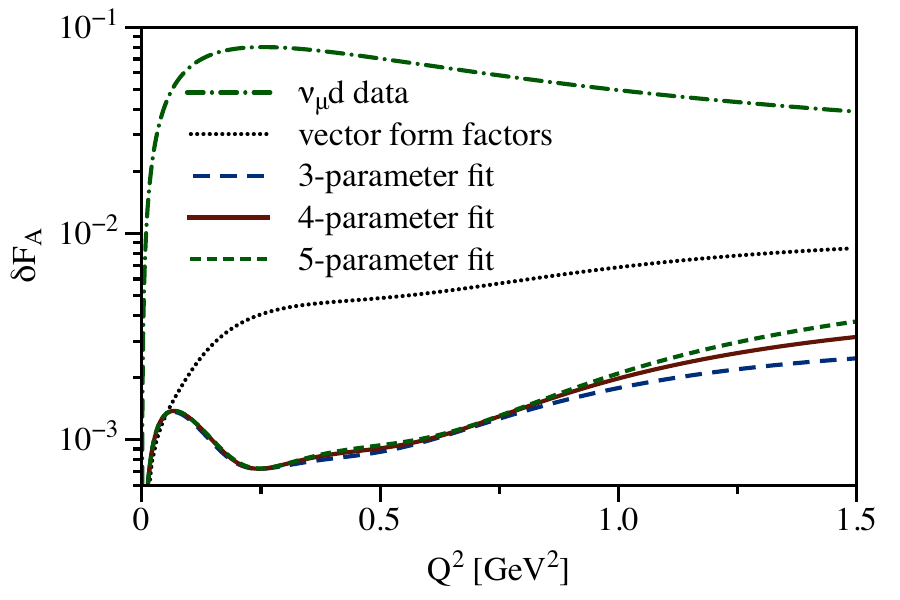} \qquad
    \caption{Projected systematic uncertainty on the nucleon axial-vector form factor obtained from antineutrino elastic scattering on hydrogen at the future DUNE/LBNF~\cite{Duyang:2019prb}, compared with the current knowledge from deuterium bubble chamber data~\cite{Meyer:2016oeg}. Data with $Q^2 < 1.5~\mathrm{GeV}^2$ are included in the fit and results are displayed for three, four, or five free parameters ($k_{\rm max}=7$, $8$, or $9$, respectively) in the $z$ expansion of $F_A$. Also shown are uncertainties from the vector form factors~\cite{Borah:2020gte}, computed with $4$ free parameters ($k_\mathrm{max} = 8$).} \label{figure:FFs}
\end{figure}

We first study the impact of the projected experimental and theoretical systematic uncertainties. To this end, we consider the experimental systematic uncertainties from Ref.~\cite{Duyang:2019prb} and initially assume infinite statistics. Figure~\ref{figure:FFs} summarizes the outcome of fits under different model assumptions. Neglecting the uncertainties on the nucleon vector form factors results in rather stringent constraints on the nucleon axial-vector form factor. We obtain comparable uncertainties on $F_A$ using Eq.~(\ref{eq:zexp}) with three, four, and five free parameters.

Once we include uncertainties from current knowledge of the vector form factors~\cite{Borah:2020gte}, we observe a significant increase in the uncertainty on the extracted values of $F_A$. As illustrated in Fig.~\ref{figure:FFs}, the error contribution from nucleon electromagnetic form factors is dominant over experimental systematics and represents an irreducible model uncertainty at present. In view of this uncertainty (with a dominant contribution from the magnetic form factor, cf. also the tension in the proton magnetic form factor after the high-precision measurements by the A1 Collaboration~\cite{Bernauer:2010wm,Bernauer:2013tpr,Ye:2017gyb,Borah:2020gte}), improved experimental and lattice-QCD determinations of the nucleon electromagnetic form factors will have an important role to play in the determination of the axial-vector form factor from future antineutrino-hydrogen elastic scattering data.

\begin{figure}[]
    \centering
    \includegraphics[height=0.4\textwidth]{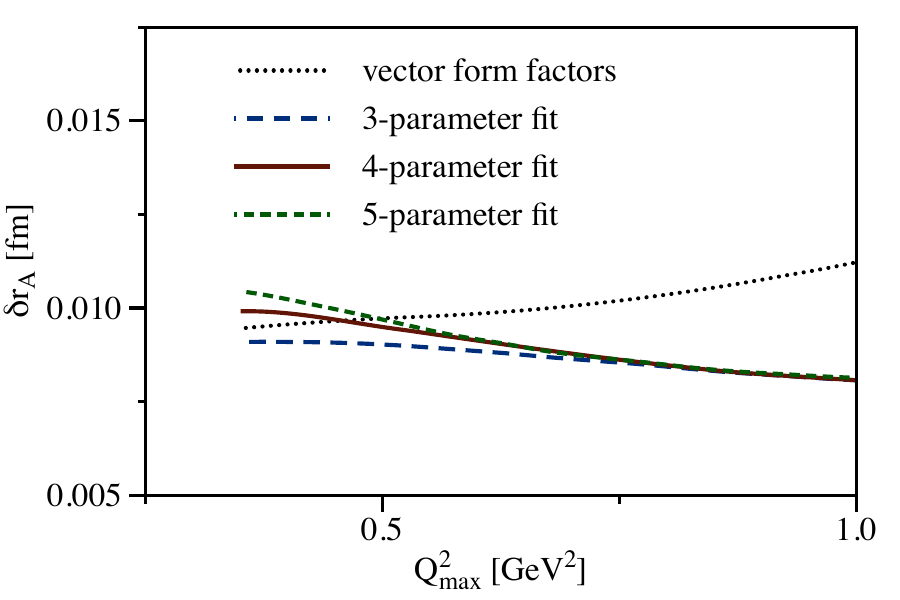}
    \caption{Uncertainty on the nucleon axial-vector radius as a function of the upper cutoff in the fitted range of the $Q^2$ distribution. The event samples are simulated using cross sections and form factors from Sec.~\ref{sec:framework}, and only systematic uncertainties are considered (infinite statistics). Results are shown for a $\chi^2$ minimization with three, four, and five free parameters ($k_{\rm max}=7$, $8$, and $9$) in the $z$ expansion of the axial-vector form factor. Also shown are uncertainties from the nucleon vector form factors, computed with four free parameters ($k_{\rm max}=8)$.} \label{figure:fig_Q2_saturation}
\end{figure}

We verify the convergence and the stability of our results by varying both the range of momentum transfer included in the fit and the number of the $z$-expansion terms considered. Figure~\ref{figure:fig_Q2_saturation} shows that the uncertainty on the nucleon axial-vector radius $r_A$ is roughly independent from the chosen $Q^2$ range once the upper cutoff is $Q^2_{\rm max}\gtrsim 0.75~\mathrm{GeV}^2$. For definiteness, we select $Q^2_\mathrm{max} = 1.0~\mathrm{GeV}^2$ in the remainder of the paper when evaluating $r_A$ uncertainties. The small difference obtained from the three-, four-, and five-parameter fits indicates an adequate convergence in the $z$-parameter series. Our results are also stable against the number of $Q^2$ bins considered: we do not find appreciable effects by increasing that number from $30$ to $60$.

We estimate the impact of radiative corrections by including the corresponding uncertainty from the hadronic model of Refs.~\cite{Tomalak:2021hec,Tomalak:2022xup} in our fit with fixed vector form factors. Although radiative corrections are significant and must be included in the analysis of current and future experimental data, we find that their uncertainty is subdominant for the determination of the axial-vector form factor using muon antineutrinos. At the projected precision in Fig.~\ref{figure:FFs}, QCD isospin-violating effects for the relation between (anti)neutrino and electron scattering should also be considered in the analysis.

\begin{figure}[]
    \centering
    \includegraphics[height=0.40\textwidth]{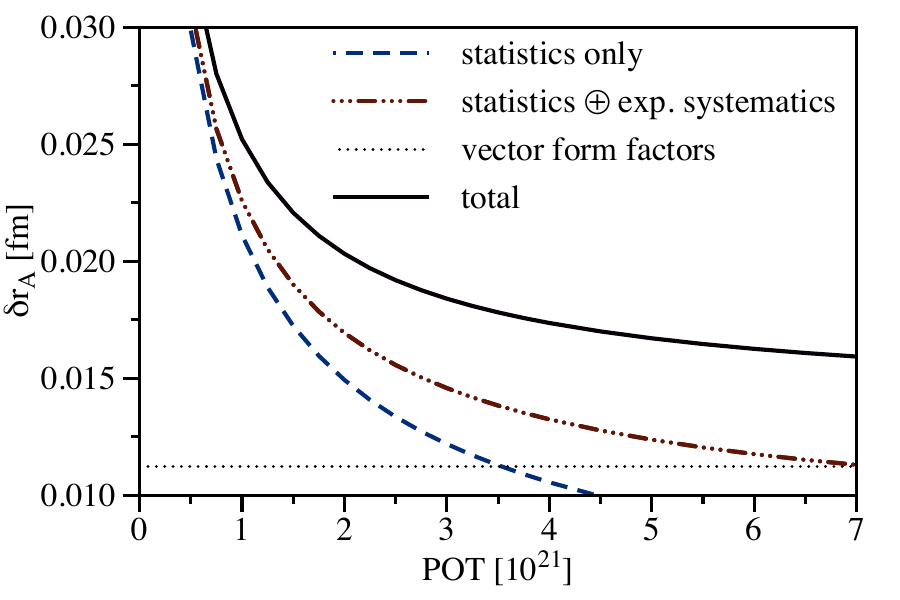}
    \caption{Total uncertainty on the nucleon axial-vector radius as a function of the LBNF beam exposure. It originates from three sources of errors. The statistical uncertainty is added in quadrature to the systematic uncertainties from the experimental measurement and from the vector form factors.} \label{figure:exposure}
\end{figure}

Having evaluated the relevant sources of systematic uncertainties, we consider the data that could be realistically collected from different exposures at the intense LBNF antineutrino beam. Figure~\ref{figure:exposure} shows the total uncertainty on the nucleon axial-vector radius, including both statistical and systematic uncertainties, achievable as a function of the exposure, expressed in terms of the number of protons on target. A comparison with the values in Fig.~\ref{figure:fig_Q2_saturation} indicates that the measurement will be statistics dominated up to exposures of about $3 \times 10^{21}$ POT. For longer exposures, theoretical and experimental systematics start to play a significant role. In the following, we will evaluate the projected results for the axial-vector form factor and radius assuming an integrated exposure of $5.5 \times 10^{21}$ POT, which is expected to be achieved in about $3.5$ years.

Table~\ref{tab:axial_radius_error} summarizes the projected uncertainty on the nucleon axial-vector radius under different assumptions. We consider separately the effect of the expected statistical and systematic uncertainties, as well as the one of their combination in quadrature. We also show the effect of including the uncertainty from the current knowledge of the vector form factors (second column).

\subsection{Fits including absolute flux normalization}

The determination of the axial-vector form factor and radius relies on the shape of the measured $Q^2$ distribution and is therefore largely insensitive to the overall normalization of the data samples. For this reason, we fixed the corresponding integral to unity in the fits described above. However, the additional information from the overall data normalization can be used to constrain the value of the absolute $\bar \nu_\mu$ flux. Since the cross section for elastic $\bar \nu_\mu p \to \mu^+ n$ process on hydrogen is fully defined by the neutron $\beta$ decay with a precision $\ll 1\%$ in the limit $Q^2 \to 0$, the integral of the incoming $\bar \nu_\mu$ flux can be directly determined from the number of events at small momentum transfer $Q^2<0.05$ GeV$^2$~\cite{Duyang:2019prb}. Rather than explicitly selecting a low-$Q^2$ sample, we extract the absolute flux simultaneously with the axial-vector form factor by expressing the overall data normalization in terms of the event rate at zero momentum transfer. Technically, we introduce the relative normalization $f$ as an additional fit parameter with its prior $\delta f$ and modify the covariance matrix~\cite{DAgostini:1993arp} by adding a common relative systematic normalization uncertainty $\delta n$ for all bins,\footnote{Fits in the Section~\ref{sec:shape_only} correspond to setting $\delta n = 0$ and $f = 1$ in Eqs.~(\ref{eq:covariance}) and~(\ref{eq:chi2}).}
\begin{align}
    \mathrm{cov}_{i j} &=  \left[ \delta_{i j} \delta r_{i} \delta r_j + \left(\delta n \right)^2 \right] N^\mathrm{gen}_i N^\mathrm{gen}_j, \label{eq:covariance} \\
    \chi^2 &= \sum \limits_{\mathrm{bins}~i,j} \left( N^\mathrm{gen}  - f N^\mathrm{fit}\right)_i \left(\mathrm{cov}^{-1} \right)_{i j} \left( N^\mathrm{gen}  - f N^\mathrm{fit} \right)_j + \sum \limits_{k=1}^{k_\mathrm{max}} \left(\frac{a_k}{5 a_0} \right)^2 + \frac{\left(f-1\right)^2}{\left(\delta f \right)^2} \,, \label{eq:chi2}
\end{align}
where $N^\mathrm{gen}_i$ and $N^\mathrm{fit}_i$ are the numbers of events generated and calculated in the $i$th bin, and $\delta r_i$ is the corresponding relative uncertainty. The dominant normalization error affecting the determination of the absolute $\bar \nu_\mu$ flux is expected to arise from the calibration of the neutron detection efficiency~\cite{Duyang:2018lpe,Duyang:2019prb}, from the detector response, and from the event reconstruction. In order to understand the constraints realistically achievable on the absolute flux and the related requirements in terms of experimental systematics, we evaluate the impact of total systematic uncertainties $\delta n$ of 1\%, 2\%, and 5\% on the overall normalization and set the prior flux uncertainty as $\delta f = 5\%$. Results are summarized in Table~\ref{tab:axial_radius_error}. As expected, the integral of the $\bar \nu_\mu$ flux can be extracted with an accuracy largely defined by the normalization systematics, while the $r_A$ determination is largely insensitive to the corresponding flux uncertainty. With normalization systematics $\lesssim$ 2\%, it seems feasible to constrain the absolute $\bar \nu_\mu$ flux below 1.9\%.

\begin{table}[p]
 \centering
 \begin{tabular}{|l|cccccc|}
 \cline{2-7}
  \multicolumn{1}{c|}{}    & ~~No VFFs~~ & With VFFs  & $\delta n = 1\%$ & $\delta n = 2\%$ & $\delta n = 5\%$ &  \\
  \multicolumn{1}{c|}{}   & error & error~\cite{Borah:2020gte} &   &   &  & \\  \hline\hline
 &  \multicolumn{5}{|c}{\bf \boldmath $\delta r_A$, fm} & \\
 Systematics only & 0.008 & 0.014 & 0.016 & 0.016 & 0.016 & \\
 Statistics only & 0.010 & 0.015 & 0.019 & 0.019 & 0.019 & \\
 Statistics $\oplus$ systematics & 0.012 & 0.016  & 0.022 & 0.022 & 0.022 & \\ \hline
 $\nu D$ (dipole)~\cite{Bodek:2007ym} & & &  & & & 0.017 \\
 $e N \to e N^\prime$ (dipole)~\cite{Bodek:2007ym} & & & & & & 0.010 \\
 Muon capture~\cite{Hill:2017wgb} & & & & & & 0.180 \\
 $\nu D$ (z expansion)~\cite{Meyer:2016oeg} & & & & & & 0.160 \\
 $\bar{\nu} H$ (z expansion)~\cite{MINERvA:2023avz} & & & & & & 0.170 \\
 RQCD (LQCD)~\cite{RQCD:2019jai} & & & & & & 0.065 \\
 ETMC (LQCD)~\cite{Alexandrou:2020okk} & & & & & & 0.038 \\
 NME (LQCD)~\cite{Park:2021ypf} & & & & & & 0.047\\
 Mainz (LQCD)~\cite{Djukanovic:2022wru} & & & & & & 0.053 \\
 PNDME (LQCD)~\cite{Jang:2023zts} & & & & & & 0.049 \\ \hline\hline
 &  \multicolumn{5}{|c}{\bf \boldmath $\delta f$, ~\%} & \\
 Statistics $\oplus$ systematics &  & & 1.1 & 1.9 & 3.5  &  \\ \hline
 \end{tabular}
 \caption{Projected uncertainties on the nucleon axial-vector radius from antineutrino elastic scattering on hydrogen at the future DUNE/LBNF with a beam exposure of $5.5 \times 10^{21}$ POT and various combinations of statistical and systematic uncertainties. Results are obtained with $Q^2_\mathrm{max} = 1.0~\mathrm{GeV}^2$ and $k_\mathrm{max} = 8$ in all determinations. Uncertainties are compared with existing measurements, as well as with recent lattice-QCD determinations (LQCD). The constraints simultaneously extracted on the absolute $\bar \nu_\mu$ flux normalization with prior flux uncertainty $5\%$ are also shown. The corresponding values are found to be independent from the bin-to-bin uncertainties considered.} \label{tab:axial_radius_error}
\end{table}
\begin{figure}[p]
    \centering
    \includegraphics[height=0.38\textwidth]{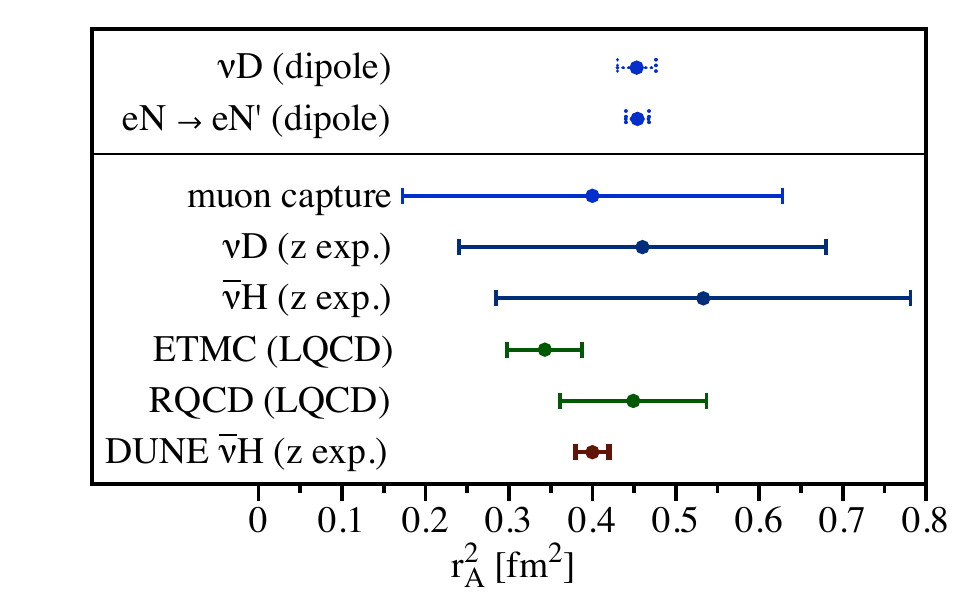}
    \caption{Comparison of existing measurements of the nucleon axial-vector radius squared with the least uncertain~\cite{Alexandrou:2020okk} and most uncertain~\cite{RQCD:2019jai} lattice-QCD estimates and the projected sensitivity at the future DUNE/LBNF for $5.5 \times 10^{21}$ POT. The latter is displayed with an arbitrary central value $r_A^2 = 0.4~\mathrm{fm}^2$. For completeness, we also include in the top part the results obtained from fits with the dipole functional form, which is known to underestimate the uncertainties.} \label{figure:axial_radius_error}
\end{figure}

\begin{figure}[]
    \centering
    \includegraphics[height=0.285\textwidth]{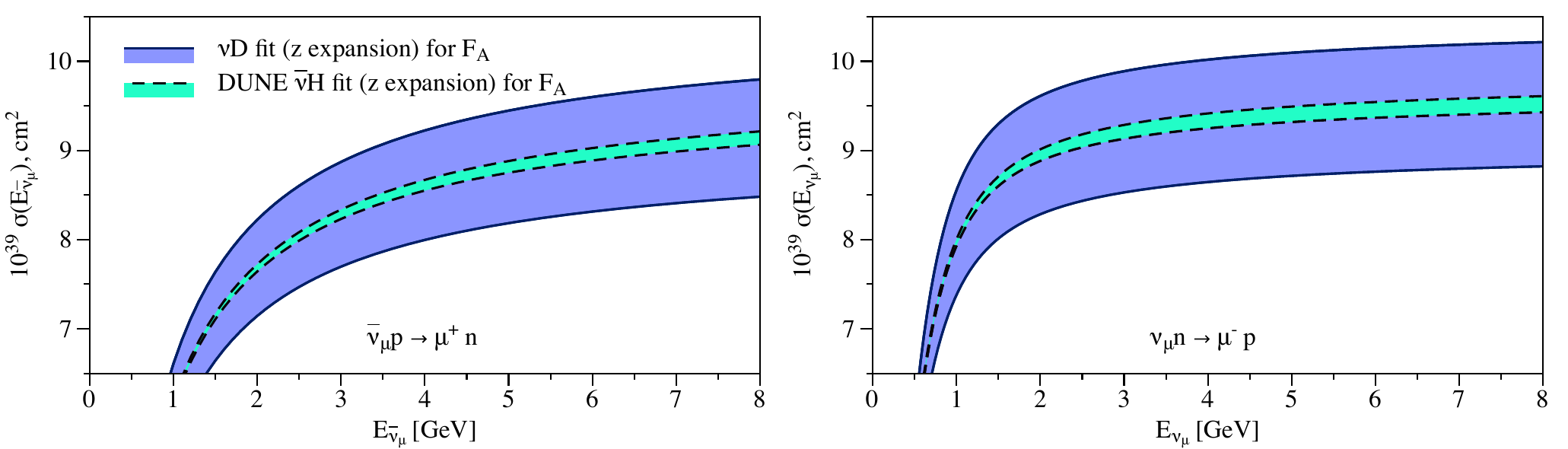}
    \caption{Projected uncertainty on neutrino-nucleon cross sections with DUNE/LBNF for $5.5 \times 10^{21}$ POT, including statistical, systematic, and vector form factor uncertainty, compared with current cross-section knowledge.} \label{figure:xsections}
\end{figure}

In Table~\ref{tab:axial_radius_error}, we compare our projected uncertainty for $r_A$ with other available determinations from neutrino-deuterium~\cite{Meyer:2016oeg,Bodek:2007ym}, antineutrino-hydrogen~\cite{MINERvA:2023avz}, muon capture~\cite{Hill:2017wgb}, and pion electroproduction~\cite{Bodek:2007ym} data, as well as from lattice QCD~\cite{Djukanovic:2022wru,RQCD:2019jai,Park:2021ypf,Alexandrou:2020okk,Jang:2023zts}. In the latter case, we have combined the various sources of uncertainties in quadrature. Although for completeness we also include older results based on the dipole functional form~\cite{Bodek:2007ym}, the $z$ expansion is generally considered to provide more realistic uncertainties by reducing the functional bias. Figure~\ref{figure:axial_radius_error} summarizes the comparisons among the various determinations of the nucleon axial-vector radius. Figure~\ref{figure:xsections} shows the projected total uncertainties for the (anti)neutrino charged-current elastic cross section on free nucleons, compared with current cross-section knowledge as given by the form factors in Sec.~\ref{sec:framework}.

\section{Summary and Outlook}

The next generation of accelerator-based neutrino oscillation experiments will put strong constraints on the electroweak structure of nucleons and nuclei. New precision measurements of antineutrino elastic scattering on hydrogen can significantly improve our understanding of the axial-vector contributions, which are currently less well known than the vector ones. In particular, high-statistics data are expected to be collected in DUNE using the solid hydrogen technique, with improved experimental systematics and a more accurate knowledge of the (anti)neutrino fluxes. With such improved neutrino scattering data, we find that the vector form factors can become a leading source of systematic uncertainty in the determination of the nucleon axial-vector form factor, with a dominant error from the magnetic form factor $G_M^V$. Using modern constraints from electron-proton scattering data, muonic hydrogen spectroscopy, and neutron radius determinations, the nucleon axial-vector radius can be extracted with a precision of $\sim 0.016~\mathrm{fm}$ from the shape of the measured $Q^2$ distribution. The absolute $\bar \nu_\mu$ flux can also be simultaneously constrained from the overall normalization in the limit $Q^2 \to 0$.

The accuracy of future measurements could constrain not only the axial-vector but also the vector form factors of the nucleon. To this end, combined fits of antineutrino-hydrogen, electron-proton, and electron-deuteron scattering data would allow an improved knowledge of the vector form factors, as well as a precision test of the isospin symmetry relations between electromagnetic and electroweak interactions. However, future precision studies will require consideration of subleading effects contributing to antineutrino-hydrogen scattering at the level of the projected uncertainties. QED radiative corrections~\cite{Tomalak:2021hec,Tomalak:2022xup} are expected to be significant and must be applied to the experimental data; the model uncertainties of radiative corrections are found to be negligible in the determination of the nucleon axial-vector radius. QCD isospin-violating effects~\cite{Dmitrasinovic:1995jt,Lewis:2006vve,Kubis:2006cy,Miller:1997ya} for the relation between (anti)neutrino and electron scattering form factors should be further investigated and included in the analyses.

Future precision measurements of antineutrino elastic scattering on hydrogen in DUNE can improve the current understanding of the unpolarized single-nucleon cross section by an order of magnitude, cf. Fig.~\ref{figure:xsections}. As summarized in Fig.~\ref{figure:axial_radius_error}, DUNE will be able to probe the nucleon axial-vector radius with an order of magnitude better precision compared to previous experimental determinations. Interestingly, the projected accuracy will be below the one obtained from recent lattice-QCD studies, offering an opportunity to resolve the reported tension~\cite{Meyer:2022mix} with respect to (anti)neutrino scattering measurements. The projected accuracy will also allow a test of the isospin symmetry of the nucleon vector form factors~\cite{Kubis:2006cy} relating electron scattering data to the (anti)neutrino charged-current process and will allow new constraints to be placed on physics beyond the Standard Model~\cite{Alvarez-Ruso:2022ctb}.

\FloatBarrier

\section*{Acknowledgments}

OT acknowledges useful discussions with Aaron Meyer, Kevin McFarland, and Clarence Wret. Research of RJH supported by the U.S. Department of Energy, Office of Science, Office of High Energy Physics, under Award Number DE-SC0019095. This work is supported by the US Department of Energy through the Los Alamos National Laboratory. Los Alamos National Laboratory is operated by Triad National Security, LLC, for the National Nuclear Security Administration of U.S. Department of Energy (Contract No. 89233218CNA000001). This research is funded by LANL’s Laboratory Directed Research and Development (LDRD/PRD) program under projects 20210968PRD4, 20210190ER, and 20240127ER. RP thanks the support from the CERN neutrino platform. This manuscript has been authored by Fermi Research Alliance, LLC under Contract No. DE-AC02-07CH11359 with the U.S. Department of Energy, Office of Science, Office of High Energy Physics. For facilitating portions of this research, OT wishes to acknowledge the Center for Theoretical Underground Physics and Related Areas (CETUP*), The Institute for Underground Science at Sanford Underground Research Facility (SURF), and the South Dakota Science and Technology Authority for hospitality and financial support, as well as for providing a stimulating environment. Mathematica~\cite{Mathematica}, Minuit~\cite{James:1975dr}, and DataGraph~\cite{JSSv047s02} were extremely useful in this work.

\bibliography{axial_radii}{}

\end{document}